\documentclass[journal]{IEEEtran}

\usepackage{ifpdf}

\usepackage{cite}

\ifCLASSINFOpdf
 \usepackage[pdftex]{graphicx}
\else
\fi
\usepackage{grffile}

\usepackage[cmex10]{amsmath}
\usepackage{algorithmic}

\usepackage{array}
\usepackage{fixltx2e}
\usepackage{url}

\newcommand{\squeezeup}{\vspace{-2.5mm}}

\begin{document}
%
\title{A Contraction Theory Approach for Analysis of Performance Recovery in Dynamic Surface Control }
%
%
%

\author{M. M. Rayguru,
        and~I. N. Kar,~\IEEEmembership{Senior~Member~IEEE}
\thanks{M. M. Rayguru is with the Department
of Electrical  Engineering, Indian Institute of Technology, Delhi,
India,  e-mail: (mmrayguru87@gmail.com).}
\thanks{I. N. Kar is working as a Professor with the Department
of Electrical  Engineering, Indian Institute of Technology, Delhi,
India,  e-mail: (ink@ee.iitd.ac.in).}
\thanks{}}

%
%

\markboth{}%
{Shell \MakeLowercase{\textit{et al.}}: Bare Demo of IEEEtran.cls for Journals}
%



\maketitle

\begin{abstract}
 Dynamic surface control (DSC) method uses high gain filters to avoid the  "explosion of complexity" issue inherent in backstepping based controller designs. As a result, the closed loop system and filter dynamics possess time scale separation between them. This paper attempts to design a novel disturbance observer based dynamic surface controller using contraction framework. In doing so the steady state error bounds are obtained in terms of design parameters which are  exploited to tune the closed loop system performance. The results not only show that DSC technique recover the performance of a backstepping controller for a small range of filter parameter but also derive the maximum bound for it. Furthermore the stability bounds are also derived in the presence of disturbances and convergence of trajectories to a small penultimate bound is proved. The convergence results are shown to hold for less conservative choice of filter parameter and observer gain. The effectiveness of the proposed controller is verified through simulation example.
\end{abstract}


%
\IEEEpeerreviewmaketitle

\section{Introduction}
%
%
%
%
\IEEEPARstart{D}{esigning} backstepping controller for nonlinear system in parametric strict feedback forms involves calculating multiple numbers of virtual control variables ($\alpha_i$) and its derivatives \cite{krstic1995,jouffroy12002,zamani2011}. The computations of these derivatives become cumbersome and unreliable with increase in order of the system. This problem is popularly termed as explosion of complexity. To overcome this problem a new methodology called dynamic surface control was introduced in \cite{swaroop2000}. The core concept of dynamic surface control (DSC) is to make the virtual control variables to pass through high gain filters thereby avoiding the differentiations involved \cite{pan2015,swaroop2000,song2004,wu2014}. Moreover the differentiability assumption on the dynamics is somewhat relaxed and the technique is also able to deal with plant uncertainties.\\
Many variants of DSC are proposed in literature depending on the uncertainties involved in system dynamics and practical implementation of the technique. An observer based DSC is proposed in \cite{song2004} using a Lyapunov based LMI approach, a neural network based DSC is proposed in \cite {wangd2005} and a DSC frameowrk robust to unknown deadzone is proposed in \cite{zhang2008}. Some of the interesting application of DSC can be found in \cite{jaga2011, zhou2014,huang2015} and references there in. However all the existing literatures utilize a high gain filter to overcome the analytical calculation of derivatives which is central concept of DSC. Recently a singular perturbation based analysis is provided in \cite{pan2015} for DSC which elaborately discuss the time scale separation in it. But the semi-global results obtained in \cite{pan2015,saberi1984} are valid for a small range of perturbation (filter) parameter and it points out that the steady state behavior of the closed loop system does not depend on selection of large control gains. \\
Relation between closed loop system response and selection of controller parameters is the most critical issue in DSC based controller designs. In this paper we propose contraction theory based approach to answer this question and quantify the stability bounds in terms of design parameters. The overall control scheme proposed here comprises of a standard DSC law along with a high gain disturbance observer. The resulting closed loop system is shown to be in three time scale. A Composite Lyapunov function based stability analysis can be carried out to find a conservative bound on the filter parameter and observer gain. However for this purpose many interconnection conditions \cite{saberi1984} need to be satisfied which are not easy to search in presence of disturbances. Instead of this approach, we adopt contraction theory which does not require any interconnection conditions and relax the conservative bound on filter parameter.  We exploit the partial contraction tools and robustness property of contractive system to derive exact relationship between design parameters and closed loop system performance which is helpful in tuning the controller. Moreover the stability bounds obtained here does not limit the filter parameter to be within a small range. The advantage of using contraction framework is more evident in presence of disturbances where arbitrarily reducing the magnitude of filter parameter and increasing the observer gain may lead to peaking phenomenon. The results shows that by proper selection of controller gains, the steady state error bounds can be reduced even for less conservative value of observer gain and filter parameter. \\ Throughout this paper, we adopt the following notations and symbols. $B_x, B_z$ denote compact subsets, $R^m$ denotes a m-dimensional real vector space. For real vectors $v$, $||v||$ denotes the Euclidean norm and for real matrix $||E||$ denotes induced matrix norm. The matrix measure induced by infinity norm is denoted as $\rho_\infty$. A metric $\Theta$ denotes a symmetric positive definite matrix and $I_n$ is an $n \times n$ identity matrix.
\subsection{Prerequisites From Contraction Theory}
Contraction is a stability tool to study differential behavior of dynamical systems \cite{lohmiller1998,forni2014,serpen2015}. A system of the form $\dot{x}=f(x,t)$ is said to be contracting if all trajectories starting inside some region in state space will converge to each other \cite{aylward2008,manchester2014}.  A region in the state space is called a contraction region if the following inequality is satisfied for the system dynamics.
\begin{equation}\label{cont1}
F=(\dot{\Theta}+\Theta\frac{\partial{f}}{\partial{x}})\Theta^{-1} \leq -\lambda I
\end{equation}
\begin{equation}\label{econt}
\Rightarrow (\dot{M}+M\frac{\partial{f}}{\partial{x}}+\frac{\partial{f}}{\partial{x}}^{T}M) \leq -2\lambda M
\end{equation}
where $\Theta$ is a nonsingular matrix, $\lambda$ is referred as contraction rate and $M=\Theta^T\Theta$ is called contraction metric. The quantity $F$ is called generalized Jacobian which should be negative definite for \eqref{cont1} to be satisfied. The quantity $-\lambda$ refers to the presence of a maximum negative eigen value or existence of a uniformly negative matrix measure \cite{sontag2014,delv2013}. Some important results  from previous literatures which are used in this paper are outlined in the form of following lemmas. Detailed proofs of these results can be found in \cite{aylward2008,lohmiller1998,delv2013,sontag2010}\par
\emph{Lemma-1}: Suppose an autonomous system $\dot{\it \bf x}=\it \bf f(\it \bf x)$ is globally contracting with a nonsingular metric $\rm \bf \Theta(\it \bf x)$ then all the trajectories of this system will converge to an unique equilibrium point.  \\
The robustness property of contraction in a perturbed nonlinear system is summarized in the form of a lemma. \par
\emph{Lemma-2}: Define a perturbed system of the following form.
\begin{equation} \label{pertsys}
\dot{\it \bf x_p}= \it \bf f(\it \bf x_p, t)+\it \bf d(\it \bf x_p,w, t)
\end{equation} 
where $w$ is external parameter. Suppose the system $\dot{\it \bf x}=\it \bf f(\it \bf x, t)$ is contracting using a nonsingular metric $\rm \bf \Theta$ with a rate $\it \lambda$. Then the following two cases arise.\\
a) Suppose the Jacobian of the perturbation term $\||\frac{\partial{\it \bf d(\it \bf x_p, w,  t)}}{\partial{\it \bf x_p}}\|| \leq \it \lambda$ for $\forall t > 0$, then the perturbed system is still contracting and the trajectories of the perturbed system will exponentially converge to the trajectories of nominal(unperturbed) system. i.e\\
\begin{equation}\label{lem21}
\lim_{t \to \infty} ||\it \bf x_p(t)-x(t)|| \rightarrow 0
\end{equation}
b) If the perturbation term is bounded in any norm, then the difference between trajectories of the perturbed system and the nominal system will converge to a steady state bound given as: 
\begin{equation}\label{lem22}
\lim_{t \to \infty} ||\it \bf x_p(t)-x(t)|| \leq \frac{\it \Upsilon \it d}{\it \lambda}
\end{equation}
where $\it \Upsilon$ is the condition number of the metric $\rm \bf \Theta$ and $||\it \bf d(\it \bf x_p, w, t)|| \leq \it d$.
\\
Apart from these properties, contraction framework provides a very useful tool called partial contraction \cite{wang2005}, which find application in filter design and synchronization problems \cite{wang2005,sharma2008}.\\
\emph{Definition-1}: Consider the dynamical system in \eqref{pertsys}. Define a auxiliary/copy system \begin{equation}\label{defpar}\dot{y}=f(y, t)+d(x_p, w, t).\end{equation} Replacing the variable $y$ in \eqref{defpar} by $x_p$ we retrieve \eqref{pertsys}. So the variable $x_p(t)$ of \eqref{pertsys} is a particular solution of \eqref{defpar}. Hence \eqref{defpar} will be referred as a virtual system for \eqref{pertsys}. \\
\emph{Lemma-3}:
A System $\dot{\it \bf x}=\it \bf f(\it \bf x, \it \bf y, t)$ is said to be partially contracting in $\it \bf x$ if an auxiliary system defined by $\dot{\it \bf z}=f(\it \bf z, \it \bf y,t)$ is contracting for any value of $\it \bf y, \forall t>0$. If the auxiliary system verifies a smooth specific property, then the trajectories of original system will also verify that property exponentially. 
\section{System Description}
We plan to discuss dynamic surface control for the systems in the following form. \\ 
\begin{equation}\label{rec1}
\begin{split}
& \dot{x}_1 = f_1(x_1)+b_1(x_1)x_2+d_1\\
& \dot{x}_2 =  f_2(x_1,x_2)+b_2(x_1,x_2)x_3+d_2\\
& \dots\\
& \dot{x}_n = f_n(x_1,x_2,x_3....x_n)+b_n(x_1,x_2,\hdots,x_n)u+d_n\\
\end{split}
\end{equation}
where $f(x)=\begin{bmatrix}f_1 & f_2 &\hdots &f_n\end{bmatrix}^T\in R^n$ is a smooth vector field, $b_i(.) \geq 0  \ \forall x \in R^n$ and $u \in R$ is the control input. The unknown disturbances $d_i$ are assumed to be slowly varying.\\ The problem is to design a control law such that the output $x_1(t)$ tracks the desired signal $x_d(t)$. \\
\emph{Assumption 1}: The desired signal $x_d(t)$ and all its derivatives are continuous and bounded.\\
\emph{Assumption 2}: The disturbance dynamics is assumed to be in the form of:
\begin{equation}\label{ddy1}
\dot{d}_i=\Xi(d_i)
\end{equation}
$\Xi(.)$ is unknown but bounded, i.e. $|\Xi(d_i)| \leq c_1$ in the region of interest where $c_1$ is a positive constant. The assumption is not restrictive in nature and satisfied for wide range of disturbances like friction, saturation etc \cite{won2015,wu2014}.

\squeezeup
\squeezeup
\subsection{Backstepping Design} 
Integrator backstepping design provides an elegant method for stabilization of strict feedback and parametric strict feedback nonlinear systems \cite{krstic1995}. The controller design is recursive in nature and can deal with uncertainties in system dynamics. We briefly summarize the methodology of backstepping controller design in contraction framework in the following lemma for \eqref{rec1} in the absence of disturbances. The detailed proof can be found in \cite{jouffroy12002,zamani2011}. \\
\emph{Lemma-4}:  In the absence of any disturbances in \eqref{rec1}, suppose a control law is selected recursively as
\squeezeup
\begin{equation}\label{bcs1}
\begin{split}
&u=\frac{1}{g(x)}[-f_n(x)-b_{n-1}z_{n-1}-\chi_n(z_n)
+ \dot{\alpha}_{(n)}
]\\
&z_i=x_{i}-\alpha_{i}, (i=1,2......m)\\
& \alpha_{1}=x_d(t) \ \text{and} \ \text{for}\  i\in [2 \hdots n]\\
&\alpha_{i}=\frac{1}{b_{i-1}}[-f_{i-1}(x_1,..x_{i-1})-\chi_{i-1}(z_{i-1})\\
& \qquad \qquad \qquad -b_{i-2}z_{i-2}
+\dot{\alpha}_{(i-1)}]\\
& \text{ $\chi_i(z_i)$  are smooth functions for which $\frac{\partial{\chi_i(z_i)}}{\partial{z_i}} > 0$.}
\end{split}
\end{equation}
\squeezeup
Then the closed loop system is in the following form.
\begin{equation}\label{sysbrs}
\begin{split}
& \dot{z}_1=-\chi_1(z_1)+b_1z_2\\
& \dot{z}_i=-b_{i-1}z_{i-1}-\chi_i(z_i)+b_{i}z_{i+1},\ \text{for}\ i \in[2, n-1]\\
& \dot{z}_n=-b_{n-1}z_{n-1}-\chi_n(z_n)
\end{split}
\end{equation}   
With the proper selection of $\chi_i(.)$,  \eqref{sysbrs} is contracting in an identity metric as its Jacobian is uniformly negative definite.$\diamond$\\
From \eqref{bcs1}, it is observed that, computation of the virtual control inputs analytically become cumbersome with the increase in the order of system. However the dynamic surface control which will be discussed in subsequent sections does not encounter this problem.
\section{Dynamic Surface Control}
 In this approach the derivative of virtual control inputs $\alpha_i(.)$ are replaced by their filtered counterparts to realize the control law  \eqref{bcs1}. The control law is given by
 \begin{equation}\label{fbcont2}
\begin{split}
&u=\frac{1}{g(x)}[-f_n(x)-b_{n-1}z_{n-1}-\chi_n(z_n)
+ \dot{\alpha}_{(n)f}-\hat{d}_n
]\\
&z_i=x_{i}-\alpha_{if}, (i=1,2......m)\\
& \alpha_{1f}=\alpha_1=x_d(t) \ \text{ and for}\  i\in [2 \hdots n]\\
&\alpha_{i}=\frac{1}{b_{i-1}}[-f_{i-1}(x_1,..x_{i-1})-\chi_{i-1}(z_{i-1})\\
& \qquad \qquad \qquad -b_{i-2}z_{i-2}
+\dot{\alpha}_{(i-1)f}-\hat{d}_{i-1}
]\\
\end{split}
\end{equation}
The scalar $\hat{d}_i$ is the estimate for $d_i$ which is obtained using a suitable observer. The signals $\alpha_{if}$ are obtained through a first order filter and expressed as:
\begin{equation}\label{fil1}
\mu\dot{\alpha}_{if}=-\alpha_{if}+\alpha_i \quad i\in[2,\hdots,n]
\end{equation}
where the initial condition $(\alpha_{if}(0)=\alpha_i(0))$ and $\mu\in[0,1]$ is a filter parameter. The closed loop system can be expressed as following form.
\begin{equation}\label{sysb}
\begin{split}
& \dot{z}_1=-\chi_1(z_1)+b_1z_2+b_1\tilde{\alpha}_2+\tilde{d}_1\\
&\dot{z}_i=b_{i-1}z_{i-1}-\chi_i(z_i)+b_{i}z_{i+1}+b_i\tilde{\alpha}_{i+1},+\tilde{d}_i\\
& \dot{z}_n=-b_{n-1}z_{n-1}-\chi_n(z_n)+\tilde{d}_n
\end{split}
\end{equation} 
where $\tilde{\alpha}_i=\alpha_{if}-\alpha_i$ and $\tilde{d}_i=d_i-\hat{d}_i$ for $i\in[2, \hdots, n-1]$.
\subsection{High Gain Disturbance Observer}
The observer structure for the estimation of disturbances in \eqref{rec1} is given by
\begin{equation}\label{hgdo1}
\begin{split}
&\dot{\xi}_i=-k(\xi_i+kx_i-f_i(x_1,x_2,\hdots,x_i)-b_ix_{i+1}) \\
&\dot{\xi}_n=-k(\xi_n+kx_n-f_n(x_1,x_2,\hdots,x_n)-b_nu) \\
&\hat{d}_i=\xi_i+kx_i\\
&\hat{d}_n=\xi_n+kx_n
\end{split}
\end{equation}
where $i \in [1,2,\hdots,n-1]$, $k >> 0$ is a positive constant, $\hat{d}_i$ is the estimate of $d_i$ and $\xi_i$ is an intermediate variable. The dynamics of disturbance estimate  is given by
\begin{equation}\label{hgdo2}
\dot{\hat{d}}_i=h(\hat{d}_i,d_i)=-k(\hat{d}_i-d_i)
\end{equation}
For a given $d_i$, denote ${d}_{ids}$ as the root of $h(\hat{d}_{ids},d_i)=0$ and define a virtual system 
\begin{equation}\label{hgvs1}
\dot{{d}}_{ids}=h({d}_{ids},d_i)+\Xi(d_i)
\end{equation}
 The  dynamics \eqref{hgvs1} can be regarded as a perturbed copy of $\hat{d}_i$.  Moreover it is partially contracting in ${d}_{ids}$ in a metric $\Theta_d=I$ with a rate $\lambda_d=k$. As the perturbation term $\Xi(d_i)$ is bounded,  we can use lemma 2 to  obtain the following inequality.   
\[\lim_{t \to \infty}||\hat{d}_i-d_{ids}|| \leq \frac{c_1}{k}\]
Hence the disturbance observer proposed in \eqref{hgdo1} will converge to a small bound depending on the gain constant $k$.\\
\textbf{Remark 1:} In case of slowly varying disturbances for which $\dot{d}_i \approx 0$, the scalar $c_1=0$.  Therefore in this special case  $\hat{d}_i \rightarrow d_{ids}$.
\section{Problem Description}
The closed loop dynamics of the overall system comprising of equations \eqref{sysb}, \eqref{fil1} and \eqref{hgdo2} can be expressed as:
\begin{subequations}\label{ovsy1}
\begin{equation}\label{snp1}
\dot{{z}}=f({z},\alpha_f,\alpha,\hat{d},d)
\end{equation}
\begin{equation}\label{fil2}
 \mu{\dot{\alpha}_f}=g(\alpha_f,\alpha)=-\alpha_f+\alpha
\end{equation}
\begin{equation}\label{hgdo3}
\epsilon {\dot{\hat{d}}}=h(\hat{d}, d)=-\hat{d}+d
\end{equation}
\end{subequations}
where $z=[z_1, z_2, \hdots, z_n]^T$, $\alpha_f=[\alpha_{2f}, \alpha_{3f}, ..., \alpha_{nf}]^T$, $\alpha=[\alpha_{2}, \alpha_{3}, ..., \alpha_{n}]^T$, $\hat{d}=[\hat{d}_1, \hat{d}_2, \hdots, \hat{d}_n]$, $d=[d_1, d_2, \hdots, d_n]$ and $\epsilon=\frac{1}{k} \in [0, 1]$. \\
 The dynamics \eqref{ovsy1}, is in singularly perturbed form and  hence the convergence of the system trajectories depends on the selection of filter parameter $\mu$ and high gain parameter $\epsilon$. The focus of this paper is to show that, the performance of a DSC law is comparable to a backstepping controller (Performance Recovery) under proper selection of filter parameter $\mu$, control gains (tuning functions) $\chi_i(.)$ and observer gain $1/\epsilon$. Using contraction theory we will derive the steady state bounds of the closed loop system as a function of controller parameters and show that these bounds can be changed according to design goal.\\
\emph{Assumption 3:} There exist two compact sets, $B_z \subset R^n$ and $B_{\alpha} \subset R^{n-1}$ respectively for $z$ and $\alpha_f$ such that $f$ and its partail derivatives are continuous and bounded.
\subsection{Performance Recovery Without Disturbance Observer}
\textbf{Theorem 1}: If all the assumptions (1, 2 and 3) are satisfied, then the following statement is true.\\
In the absence of disturbances there exists a $\mu^* \in [0, 1]$  such that the control law \eqref{fbcont2} recovers the performance of a backstepping  controller for $\mu \in [0, \mu^*]$ inside $(B_z \times B_{\alpha})$ and it can be selected following \eqref{fils1}.\\
\textbf{Proof:} In the absence of disturbance terms, the control law \eqref{fbcont2} will not depend on $\hat{d}_i$. In this case the closed loop system can be written as
\begin{subequations}\label{ovsis1}
\begin{equation}\label{snp11} 
\dot{{z}}=f({z},\alpha_f,\alpha)
\end{equation}
\begin{equation}\label{fil21}
 \mu{\dot{\alpha}_f}=g (\alpha_f,\alpha)=-\alpha_f+\alpha
\end{equation}
\end{subequations}
where $f(.)=[f_1(.), f_2(.), \hdots, f_n(.)]^T$ and
\begin{equation*}\label{sysbad}
\begin{split}
& f_1(.)=-\chi_1(z_1)+b_1z_2+b_1\tilde{\alpha}_2\\
&f_i(.)=b_{i-1}z_{i-1}-\chi_i(z_i)+b_{i}z_{i+1}+b_i\tilde{\alpha}_{i+1},\\
& f_n(.)=-b_{n-1}z_{n-1}-\chi_n(z_n)
\end{split}
\end{equation*} 
where $\tilde{\alpha}_i=\alpha_{if}-\alpha_i$ and $i \in [2, \hdots n-1]$. For any given $\alpha$, the equation $g(\alpha_{f},\alpha)=0$ has an unique root given by $\alpha_{f}=\alpha$, which we  denote as $\alpha_{des}$. Using $\alpha_{des}$ we define a virtual system,
\begin{equation}\label{snp2}
\mu\dot{\alpha}_{des}=g  (\alpha_{des},\alpha)+\mu Q(z,\alpha,\alpha_{des})
\end{equation}
where $Q(z,\alpha_{des},\alpha)=\frac{\partial{\alpha_{des}}}{\partial{z}}\dot{z}$ and $g(.)=-\alpha_{des}+\alpha$. The auxiliary system \eqref{snp2} can be regarded as perturbed copy of \eqref{fil2}. From the structure of $g(.)$, the dynamics of \eqref{snp2} is partially contracting in $\alpha_{des}$ when the perturbation term $Q(.)$ is absent. \\
From assumption 3 the term $f(.)$ and its partial derivatives are bounded inside $(B_z \times B_{\alpha})$, so it is reasonable to assume $Q(.)$ and its partial derivatives are bounded inside $(B_z \times B_{\alpha})$. Suppose there exists a positive constant $c_2$ such that
\begin{equation}\label{infn1}
||\frac{\partial{Q(z,\alpha,\alpha_{des})}}{\partial{\alpha_{des}}}||\leq c_2 \ \text{inside} (B_z \times B_{\alpha}).
\end{equation}
 The differential dynamics of \eqref{snp2} can be written as
\begin{equation}\label{fil3}
\begin{split}
& \mu \ \delta \dot{\alpha}_{des}=\{\frac{\partial{g(\alpha_{des},\alpha)}}{\partial{\alpha_{des}}}+\mu\frac{\partial}{\partial{\alpha_{des}}}(Q(z,\alpha,\alpha_{des}))\}\delta {\alpha}_{des}\\
& \Rightarrow \mu \ \delta \dot{\alpha}_{des}=\{-I+\mu\frac{\partial}{\partial{\alpha_{des}}}(Q(z,\alpha,\alpha_{des}))\}\delta {\alpha}_{des}
\end{split}
\end{equation} Select  a small positive constant $\mu^*$ such that 
\begin{equation}\label{fils1}
\mu^*c_2 \leq 1.
\end{equation}
Using lemmas 2 and  3, the virtual system \eqref{snp2} is partially contracting in $\alpha_{des}$ with a contraction rate $\frac{1-||\mu c_2||}{\mu}$ for all $\mu \in [0, \mu^*]$. Therefore, the trajectories of system \eqref{fil2} and \eqref{snp2} will converge to each other.
\begin{equation}\label{bod1}
\lim_{t \to \infty} ||\alpha_f(t)-\alpha_{des}(t)|| \rightarrow 0
\end{equation}
From \eqref{fil2}, $\alpha_{des} = \alpha$ and therefore
\begin{equation}\label{bd1}
\lim_{t \to \infty} ||\alpha_f(t)-\alpha(t)|| = \lim_{t \to \infty} || \tilde \alpha|| \rightarrow 0 \end{equation}
 Replacing \eqref{bd1} in \eqref{ovsis1}, the reduced system obtained can be written in compact form as
 \begin{equation}\label{snp3}
\dot{\bf{z}}_r=f(\bf{z}_r,\alpha_{des},\alpha)
\end{equation}
where $z_r$ denotes the states of the reduced slow system. The reduced system is in same form that will result  from a integrator backstepping design \eqref{sysbrs} and therefore is contracting in $z_r$. 
 Lets rewrite \eqref{snp11} as
\begin{equation}\label{snp4}
\dot{\bf{z}}=f(\bf{z},\alpha_{des},\alpha)+f(\bf{z},\alpha_f,\alpha)-f(\bf{z},\alpha_{des},\alpha)
\end{equation}
This equation can also be regarded as a perturbed version of the reduced slow system \eqref{snp3}. Suppose \eqref{snp3} is contracting in a metric $\Theta_z$ with a rate $\beta$ and the function $f$ is Lipschitz in $\alpha_f$ with a constant $L_1$. The perturbation  term in \eqref{snp4} verify the following bound. 
\begin{equation}\label{snp5}
||f(\bf{z},\alpha_f,\alpha)-f(\bf{z},\alpha_{des},\alpha)|| \leq L_1||\alpha_f (t)-\alpha_{des} (t)||
\end{equation}
Using lemma 2 and \eqref{bd1}, we obtain the following result.
\begin{equation}
\lim_{t \to \infty} ||z(t)-z_r(t)|| \rightarrow 0
\end{equation}
Therefore the DSC design will recover the performance of a backstepping controller exponentially if the filter parameter $\mu$ is chosen such that $\mu \leq \mu^*$ according to theorem 1. $\diamond$\\
\textbf{Remark 2:}  The authors of \cite{pan2015} has shown the dynamics of DSC possesses two time scale property and proved the closed loop system stability inside a small range of filter parameter $\mu$. However Theorem 1 not only proves the closed loop system stability but also answers, how to select the filter parameter for full performance recovery. The bound obtained in \eqref{fils1}, provides the maximum value of filter parameter for which a DSC law will recover the performance of a backstepping controller in absence of any disturbances. A quadratic Lyapunov function \cite{saberi1984} approach can also be utilized to obtain the stability bounds, but it requires many interconnection conditions to be satisfied. 
\squeezeup
\squeezeup
\subsection{DSC With Disturbance Observer}
Presence of disturbances in \eqref{rec1}  complicates the selection of filter parameter $\mu$ which can not be made arbitrary small due to sampling requirements and noise. However it is possible to guarantee convergence of trajectories to a ultimate bound for a wider range of filter parameter and observer gain. The main result of this paper is summarized in the form of a theorem. \\
\textbf{Theorem 2:} Let the assumptions (1, 2 and 3) are satisfied for the system \eqref{rec1} and there exists a positive constant $c_3$ such that $|\frac{\partial{\alpha_{des}}}{\partial{z}}\dot{z}| \leq c_3$ inside $B_z \times B_{\alpha}$. Then the trajectories of the closed loop system \eqref{ovsy1} will follow the bounds given in \eqref{bound1} and \eqref{bound2}.\\ 
\textbf{Proof:} 
Express the fast subsystem \eqref{fil2}, \eqref{hgdo3} as:
\begin{equation}\label{fss1}
\mu \dot{v}=Q(v, \alpha, d, \kappa)
\end{equation}
where $\epsilon=\mu \kappa$, $v=[\alpha_f ^ T, \hat{d} ^T]^T$ and $Q(.)=[g^T(.),  \frac{1}{\kappa}h^T(.) ]^T.$
The Jacobian of \eqref{fss1} is given by
\[J_v=\frac{\partial{Q}}{\partial{v}}=\begin{bmatrix}J_\alpha\\J_d\end{bmatrix}=\begin{bmatrix}\frac{\partial{g(.)}}{\partial{v}}\\\frac{1}{\kappa}\frac{\partial{h(.)}}{\partial{v}}\end{bmatrix}\]
We will show that $J_v$ is negative definite and therefore \eqref{fss1} is partially contracting in $v$. From \eqref{fil1} and \eqref{fil2}, the matrix measure $\rho_{\infty}$ corresponding to infinity norm for $J_\alpha$ and $J_d$ are given by
\[\rho_\infty^\alpha=max_i(J_\alpha^{ii}+\sum_{i\neq j}||J_\alpha^{ij}||)=-1.\]
\[\rho_\infty^d=max_i(J_d^{ii}+\sum_{i\neq j}||J_d^{ij}||)=-1/\kappa.\]
It is important to note that $\alpha$ is treated as an external variable in \eqref{fil2} while evaluating $J_\alpha$.
Therefore the matrix measure $\rho_{\infty}^v$ is given by
\[\rho_{\infty}^v=max(-1, \frac{-1}{\kappa}).\]
As $\kappa$ is positive constant, $\rho_\infty^v$ is negative. Hence \eqref{fss1} is partially contracting in $v$.\\
\textbf{Note:} Depending on the magnitude of $\kappa$ which may be greater than or less than unity, the dynamics of disturbance observer is slower or faster than the filter dynamics. However we will show that it is not necessary to select a very small value of $\epsilon$ (or a very large value of observer gain) for better performance.\\
Following lemma I, there exists an unique root of the equation $Q(v, \alpha, d, \kappa)=0$, which is denoted as: \[v_{ds}=[\alpha^T,  d^T]^T\]
Define a virtual system given by:
\begin{equation}\label{virfs1}
\mu \dot{v}_{ds}=Q(v_{ds},\kappa)+\mu \dot{v}_{ds}
\end{equation}
In the absence of the perturbation $\dot{v}_{ds}$, \eqref{virfs1} is partially contracting in $v_{ds}$ in identity metric with a rate $1/\mu$. Following similar line of arguments as in theorem 1, the system \eqref{virfs1} can be regarded as perturbed virtual system of \eqref{fss1}.  Suppose there exists a positive constant $c_3$ such that 
\begin{equation}\label{infa1}
||\frac{\partial{\alpha_{}}}{\partial{z}}\dot{z}|| \leq c_3 \ \text{inside $B_z \times B_{\alpha}$.}
\end{equation}  From assumption 2 and \eqref{infa1}, the perturbation term $\dot{v}_{ds}$ is bounded inside the region $\rm B_z \times B_\alpha$ i.e,
\[||\dot{v}_{ds}|| \leq max(c_1,c_3) \] 
 Exploiting the boundedness of the perturbation term and using lemma 2, the trajectories of \eqref{fss1} satisfies the following bound.

\begin{equation}\label{bound1}
||v(t)-v_{ds}(t)|| \leq ||v(0)-v_{ds}(0)||e^{(-1/\mu)t}+ max(c_1, c_3)
\end{equation}
replacing $v_{ds}$ in \eqref{ovsy1}, the reduced system can be written as:
\begin{equation}\label{redss1}
\dot{{z}}_r=f({z}_r,v_{ds},\kappa,\alpha,d)
\end{equation}
 The system \eqref{redss1} is in same form as \eqref{sysbrs} and hence is contracting in $z_r$ with a metric $\Theta_z$.
The dynamics of \eqref{snp1} can be expressed as following virtual system.
\begin{equation}\label{virss1}
\dot{{z}}=f({z},v_{ds},\kappa,\alpha,d)+f({z}_r,v,\kappa,\alpha,d)-f({z},v_{ds},\kappa,\alpha,d)
\end{equation}
Exploiting the Lipschitz property of $f(.)$, the perturbation term in \eqref{virss1} satisfy the following inequality.
\begin{equation}\label{virss1}
||f({z}_r,v,\kappa,\alpha,d)-f({z},v_{ds},\kappa,\alpha,d)|| \leq L_v|v(t)-v_{ds}(t)|
\end{equation}
where $L_v$ is the Lipschitz constant. From the bound \eqref{bound1} and lemma 2, the trajectories of \eqref{snp1} converge to the following steady state bound.
\begin{equation}\label{bound2}
\lim_{t\rightarrow \infty}|z(t)-z_{ds}(t)| \leq \mu\frac{C_z L_v (max(c_1,c_3))}{\lambda_z ||max(-1, -1/\kappa)||}
\end{equation} 
where $C_z$ is the condition number of the metric $\Theta_z$ and $\lambda_z$ is contraction rate of the reduced slow system \eqref{redss1}.\\
\textbf{Parameter Selection:}  Theorem 2 discusses the effect of disturbances in the performance of DSC technique. It provides precise steady state error bounds for closed loop system trajectories depending on selection of observer and controller gains. The effect of disturbances can be reduced for sufficiently high value of observer gain but it may lead to peaking phenomenon inherent in high gain observers. We can use a moderate value of observer gain and still reduce the performance deficit by properly choosing the tuning functions $\chi(.)$.  One can select the tuning functions as $\chi_i(.)=k_c z_i$ with the proper choice of $k_c$ using \eqref{bound2}. Larger control gains $k_c$ leads to large contraction rate $\lambda_z$ thereby small steady state error. The bound can also be decreased by selecting a low value of filter parameter $\mu$, but it can not be selected arbitrarily small.  \\
\textbf{Remark 3:} It is shown in theorem 2 that, the disturbance observer based DSC can not fully recover the performance of a backstepping controller  in presence of rate bounded disturbances. However when the disturbance is slowly varying and can be approximated by $\dot{d} \approx 0$, full performance recovery is possible within a small range of perturbation parameter. This can be proved using the similar approach as theorem 1.

\section{Simulation Examples}
Consider an example of  D.C motor driven manipulator \cite{slotine1991}.
\begin{equation}\label{dcrm1}
\begin{split}
& \dot{x}_1=x_2\\
& \dot{x}_2=\frac{-N}{M} sin x_1 -\frac{B}{M}x_2+\frac{1}{M}x_3\\
&\dot{x}_3=-\frac{K_b}{L}x_2-\frac{R}{L}x_3+\frac{1}{L}u \\
\end{split}
\end{equation}Where the three states $x_1, x_2, x_3$ correspond to position, velocity and armature current respectively and $u$ is the input voltage. The parameters of the system \eqref{dcrm1} is given in appendix. The controller design objective is position tracking of a desired signal $x_d= (\pi/2) sin (8 \pi t/ 5)$. 
The constant parameters of this system are given by,
$K_b=0.90\ N m/A, R=5 \Omega, L=0.025 H, M=0.0640, B=0.0044, N=2.2816.$ \\
Initially the system \eqref{dcrm1} is simulated without the disturbance terms. The tuning functions are chosen as
$\chi_i(.)=-5z_i$. The expression of $\alpha_i$ are calculated analytically for comparing the performance of DSC with integrator backstepping control law.  Using \eqref{fils1} the value of $\mu^*$ obtained is given by $\mu^*=1/83$.  The simulation is done using a value of $\mu=0.01$ and initial conditions $[2\pi, 0, 0]$. The simulation results are given in figure 1.\\
\begin{figure}[h]
\centering
\includegraphics[width=3.25in,height=2.00in]{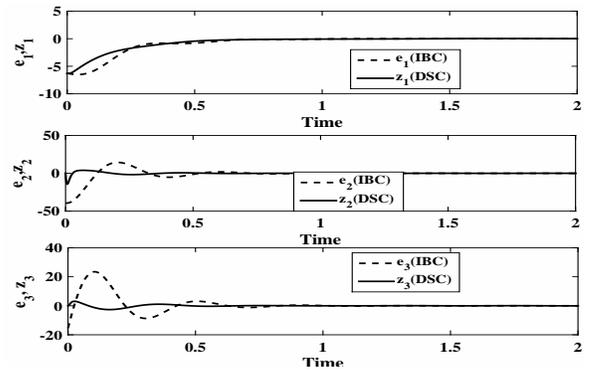}
\caption{Closed loop System Response without disturbances}
\end{figure}
The closed loop system trajectories for  backstepping controller and DSC are denoted as $[e_1, e_2, e_3]$ and $[z_1, z_2, z_3]$ respectively. From the figures 1, it is observed that DSC law recovers the performance of a backstepping based control law for the given choice of design parameters. The magnitude of filter parameter $\mu$ can be increased up to a threshold  value of $\mu^*$, but beyond that full performance recovery is difficult to achieve.\\
The simulation is also performed with disturbances  in \eqref{dcrm1}. The disturbances are the combination of both columb friction and periodic disturbances and given by
\[d_1=0.2sgn(x(2))+10sin(2t+1)+10t, d_2=10cos(2t+1)\]
  The observer gain in \eqref{hgdo1} is selected as $k=50$ and the control gains (tuning functions) are kept same as that of previous case. It is observed from figure 2 that, the steady state response for these choice of parameters are not satisfactory.
The simulation is then repeated by selecting another set of tuning functions (control gains)
 $\chi_i(.)=-40z_i$. Figure 3 confirms the improvement of response inequality, which justify the selection of $\chi_i(.)$ to reduce the steady state error bound. The procedure does not need the observer gain to be very large and hence peaking phenomenon can be avoided.
 \squeezeup
 \begin{figure}[h]
\centering
\includegraphics [width=3.25in,height=2.000in] {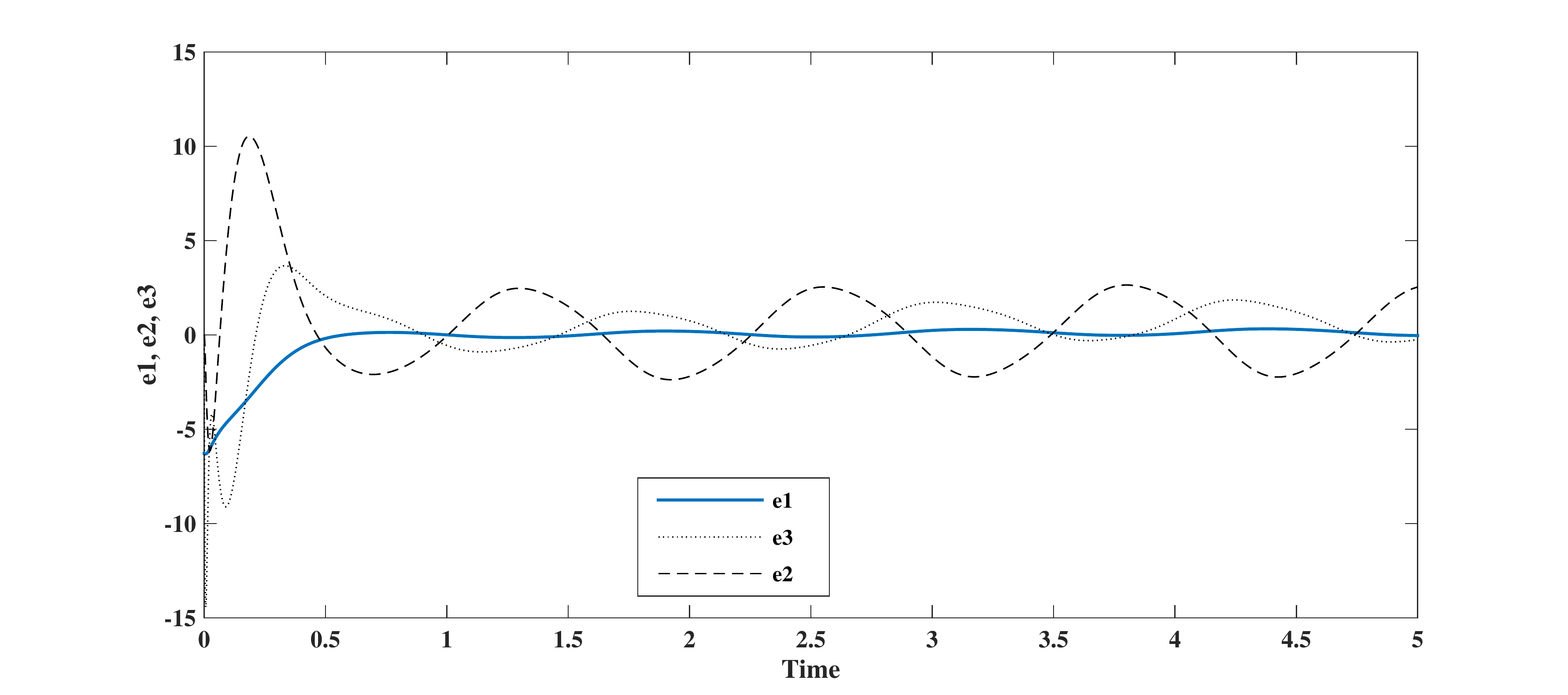}
\caption{Closed loop System Response with disturbances (low control gains) }
\end{figure}
\squeezeup
\begin{figure}[h]
\centering
\includegraphics [width=3.25in,height=2.000in] {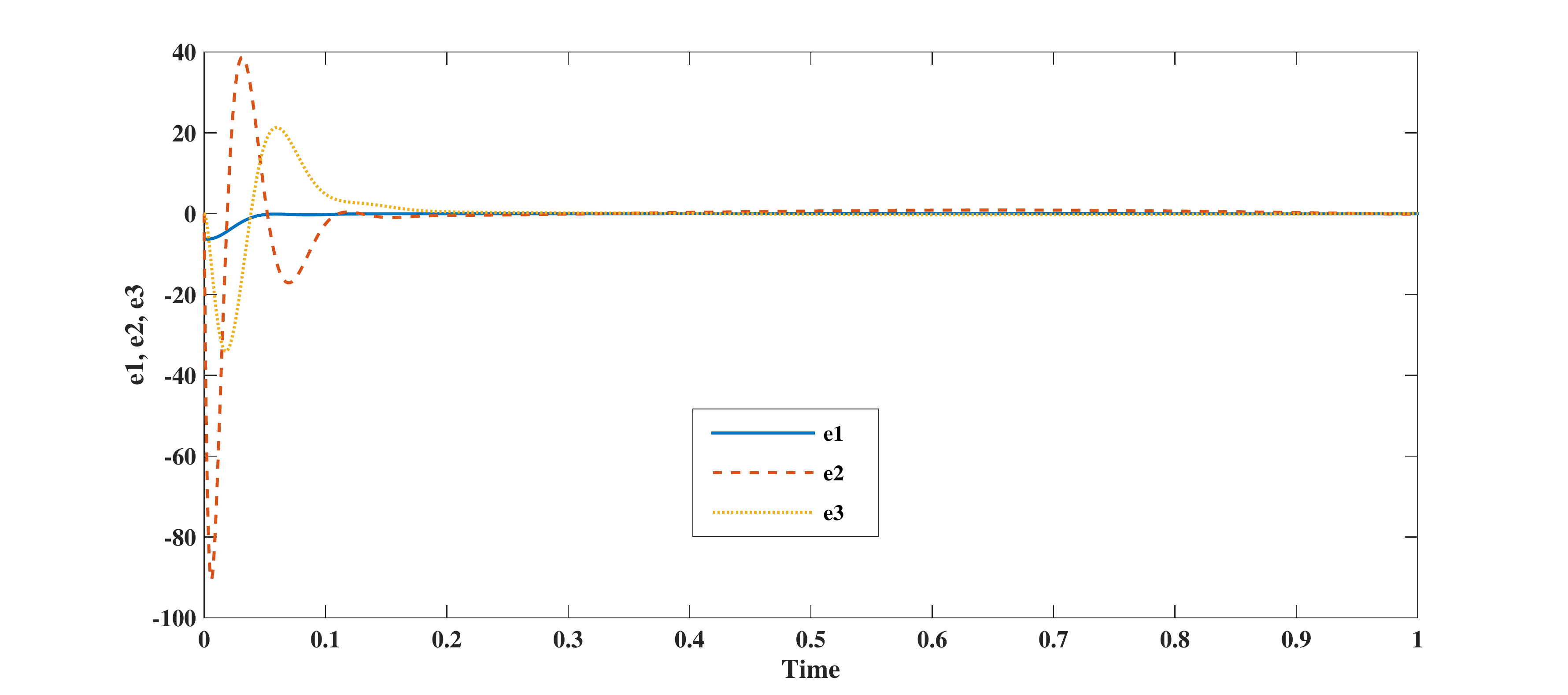}
\caption{Closed loop System Response with disturbances (high control gains)}
\end{figure}
\squeezeup

\section{Conclusion}
In this paper we proposed a new contraction theory based approach to design DSC law and tune its parameters. Steady state error bounds for DSC law are derived in terms of design parameters. The methodology presented here  relaxes the conservative bound on filter parameter which can not be made arbitrarily small due to sampling requirements and noise. Moreover it is proved that proper selection of control gains can reduce the dependency on higher observer gain for better performance of the DSC law. As a result the peaking phenomenon inherent in high gain observers can be avoided. The proposed contraction based method has important practical application in robotics, electro-hydraulic systems etc which is the subject of future works.   \\


%

\squeezeup
\squeezeup
\squeezeup




\ifCLASSOPTIONcaptionsoff
  \newpage
\fi



%

\bibliographystyle{plain}
\bibliography{tCONguide}

%







\end{document}